\documentclass[%
 reprint,
 amsmath,amssymb,
 aps,
 pra,
floatfix]{revtex4-2}
\usepackage{siunitx}
\usepackage{xcolor}
\usepackage{booktabs}
\usepackage{hyperref}
\usepackage{graphicx}
\usepackage{dcolumn}
\usepackage{bm}
\usepackage{floatrow}

\usepackage[markup=underlined,defaultcolor=red,]{changes}
\usepackage[caption=false]{subfig}
\DeclareSIUnit\gauss{G}
\DeclareSIUnit\ppm{ppm}
\DeclareSIUnit\phonon{phonon}

\newcommand{\ket}[1]{|#1\rangle}

\newcommand{\phantomsubfloat}[1]{
    {
        \captionsetup[subfloat]{farskip=0pt,captionskip=0pt}
        \captionsetup[subfigure]{labelformat=empty}
        \subfloat{#1}
    }%
}

\usepackage{comment}
\begin{document}


\title{Predicting electric-field noise in ion traps using fluctuation electrodynamics}

\author{Markus Teller$^{1,2*}$}
\author{Da An$^{3}$}
\author{Alberto M. Alonso$^{3}$}
\author{Philip C. Holz$^{4}$}
\author{Philipp Schindler$^{1}$}
\author{Hartmut Häffner$^{3,5}$}
\author{Tracy E. Northup$^{1}$}
\affiliation{$^{1}$Institut für Experimentalphysik, Universität Innsbruck, Technikerstra\ss e 25, 6020 Innsbruck, Austria}
\affiliation{$^{2}$Institut de Ciencies Fotoniques, The Barcelona Institute of Science and Technology, Castelldefels (Barcelona) 08860, Spain}
\affiliation{$^{3}$Department of Physics, University of California, Berkeley, CA 94270, USA}
\affiliation{$^{4}$Alpine Quantum Technologies GmbH, Technikerstraße 17/1, 6020 Innsbruck, Austria}
\affiliation{$^{5}$Lawrence Berkeley National Laboratory, Berkeley, CA 94720, USA}
\date{\today}

\begin{abstract}
Electric-field noise in microfabricated surface ion traps contributes to gate errors in trapped-ion quantum computers, but this noise has been challenging to predict. 
Here, we present a method to predict the electric-field noise arising from fluctuations in the bulk of dielectric and metallic materials. The method is valid for arbitrary trap geometries, and the only relevant material property is energy loss. We apply it to an ion trap with an electrically floating electrode and find the contribution from dielectrics to be dominant. 
Our model predicts the observed noise for this trap within a factor of four. In addition, we apply our model to typical surface trap designs 
and analyze how the electrode structures shield the noise.

\end{abstract}
\maketitle

\section{\label{sec:Intro}Introduction}


Quantum information processors have the potential to outperform 
classical systems for certain useful computing and communication tasks, but realizing this potential will require the leading hardware platforms to address crucial materials challenges~\cite{deleon2021materials}. 
Electric-field noise is one of these challenges. 
Here, we consider microfabricated surface ion traps, in which ions are typically located just tens of micrometers above the trap surface, such that electric-field noise is of particular concern. This noise couples to the shared motion of the ions~\cite{Turchette2000} and thus can limit the fidelity of quantum gate operations~\cite{PhysRevLett.94.153602,Monroe95a}. Most studies focus on contaminants and adsorbates on the trap electrodes as sources of surface noise~\cite{PhysRevA.84.053425,PhysRevA.99.063427,Daniilidis2014,safavinaini2013,Kumph_2016,foulon2022}. However, 
losses in the bulk --- that is, due to the electrodes and dielectric materials of the ion trap itself--- 
may also lead to significant noise~\cite{Henkel1999,Kumph_2016}.

Previously, we developed a method to predict the electric-field noise from dielectric materials and applied it to a dielectric mirror close to a single ion~\cite{Teller2021}. The method relies on the stationary solution of Maxwell's equations; consequently, time-dependent effects present in metals, such as radio-frequency shielding and skin effects, are not considered. We showed that this static analysis can predict the noise originating from mostly unshielded dielectric materials. The method of Ref.~\cite{Teller2021} has recently been applied to predict the noise from dielectrics at cryogenic environments~\cite{bruff2025}.
However, this time-independent analysis was not capable of predicting the noise due to the bulk material for an entire ion trap, which includes both metal electrodes and dielectrics. Here, we extend the static method of Ref.~\cite{Teller2021} to the time domain, allowing us to predict 
the noise due to both metals and dielectrics for an arbitrary ion-trap geometry. 

We present two examples in which the time-dependent approach enables us to analyze the effect of noise from the bulk on ion heating. First, we apply the new method to the surface ion trap described in Ref.~\cite{an2018}, in which the motion of two ions was recently coupled through an electrically floating metallic wire~\cite{an2022coupling}. Comparing our predictions to experimental data, we gain insight into the role of floating metal structures in heating the ions' motion. In contrast to existing methods, our method allows to predict electric-field noise originating from the bulk in an arbitrary ion trap geometry. As a second example, we analyze how the geometry of the gaps between electrodes affects the shielding of noise originating from the underlying dielectric substrate, an important design consideration for trap fabrication.
 We anticipate that it will be an important tool not only for analyzing existing traps but also for designing new traps.

\section{Method to calculate the electric field noise}\label{Sec:Theory}

The key concept of our method is that the fluctuation-dissipation theorem allows us to link noise to dissipation. We calculate the dissipation of the dynamic electric fields inside both dielectrics and metals due to a moving test charge; in contrast to our previous work~\cite{Teller2021}, we consider not only dielectrics but also metals. From the dissipation, we find the noise experienced by the test charge. We then convert this noise into a lower bound on the electric-field noise.

\begin{figure*}
\phantomsubfloat{\label{fig:1a}}
\phantomsubfloat{\label{fig:1b}}
\includegraphics[scale=0.27]{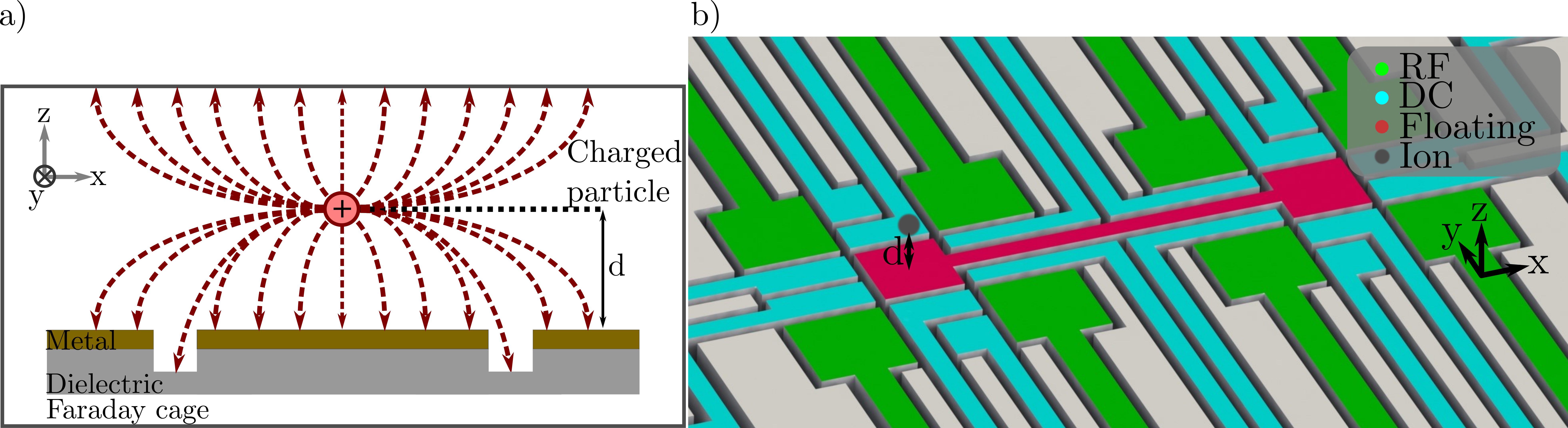}
\caption{a) A charged particle is trapped at a distance $d$ above a generic surface ion trap. Metal electrodes partially cover a dielectric substrate. b) The surface ion trap that we use~\cite{an2022coupling,an2018} includes one electrically floating metallic wire in addition to RF and DC electrodes.  A single $^{40}$Ca$^+$ ion is trapped at a distance $d$ above the wire. 
}\label{fig:figure1}
\end{figure*}

The elements of the method are illustrated in Fig.~\ref{fig:1a}. A particle with mass $m$ and charge $q$ is trapped at a distance $d$ above a surface ion trap. The trap consists of metallic electrodes on a dielectric substrate. As is common for surface ion traps, the electrodes are separated by trenches, and the dielectric in these trenches is partially exposed. Both materials are assumed to be at standard temperature, $T = \SI{293.15}{\kelvin}$. We define the $xy$ plane of our coordinate system to be parallel to the metal surface and the $z$ axis to be orthogonal to the surface, with the origin at the potential minimum. The normal modes of the particle's motion align with the three coordinate axes; the oscillation along axis $\xi\in \{x,y,z\}$ is at frequency $\omega_\xi$ with amplitude $\alpha_\xi$. The particle's charge generates an oscillating electric field at frequency $\omega_\xi$; this dynamic electric field is present in both materials. A Faraday cage, acting as a far-field ground, encloses both the metal-dielectric structure and the particle.

In the following, we refer to the materials with index $k\in\{\textrm{m},\textrm{d}\}$, where m refers to the metal and d to the dielectric. Each material is characterized by its complex permittivity $\varepsilon_k$.
The complex permittivity of the dielectric is $\varepsilon_\mathrm{d} = \varepsilon_0 \varepsilon_\mathrm{r}(1+i\tan{\delta})$, where $\varepsilon_0$ is the vacuum permittivity, $\varepsilon_\mathrm{r}$ the relative permittivity, and $\tan{\delta}$ the loss tangent. The metal layer has conductivity $\sigma$, and its complex permittivity can be approximated as $\varepsilon_\mathrm{m} = i\sigma/\omega_\xi$~\cite{Kumph_2016}. 
The electric field inside each material is damped, 
corresponding to the mean dissipated power $\bar{P}_k$~\cite{Kumph_2016}. The fluctuation-dissipation theorem links $\bar{P}_k$ to the power spectral density of the electric-field noise
\begin{equation}
	S_{\mathrm{E},k}(\omega_\xi)=\frac{8k_\mathrm{B}T \bar{P}_k}{(q \alpha_\xi \omega_\xi)^2} \mathrm{,} \label{eq:noise}
\end{equation}
at the particle's position $\mathbf{r} = (x,y,z)$, as can be seen by combining Eq.~3 and Eq.~11 of Ref.~\cite{Kumph_2016}. Here, $k_\mathrm{B}$ corresponds to the Boltzmann constant.

The electric-field noise can thus be determined from the dissipated power in the materials, 
which we calculate using 
finite-element analysis (FEA) software~\footnote{COMSOL Multiphysics 5.6}. As the particle's motional amplitude is much smaller than $d$, we model the oscillating particle in software as an electric point dipole with dipole moment vector
 \begin{equation}
\mathbf{p}_{\textrm{c},\xi} = q\omega_\xi \alpha_\xi \hat{\mathbf{u}}_\xi
\end{equation}
pointing in the direction of the unit vector $\hat{\mathbf{u}}_\xi\in\{\hat{\mathbf{x}},\hat{\mathbf{y}},\hat{\mathbf{z}}\}$. We numerically solve Maxwell's equations in the frequency domain for $\omega_\xi$, determine $\bar{P}_\mathrm{m}$ and $\bar{P}_\mathrm{d}$, and evaluate Eq.~\ref{eq:noise} for both materials with the MUMPS solver~\footnote{\url{https://mumps-solver.org}}. The total electric-field noise is then the sum of $S_\mathrm{E,m}$ and $S_\mathrm{E,d}$, based on the assumption that the noise terms are uncorrelated. 

In trapped-ion experiments, electric-field noise manifests itself in the motional heating of ions. 
The heating rate $\dot{\bar{n}}_\xi$ is related to the noise spectral density by~\cite{Brownnutt2015}\begin{equation}
\dot{\bar{n}}_\xi(\omega_\xi) = \frac{q^2}{4m\hbar\omega_\xi}S_\mathrm{E}(\omega_\xi)\mathrm{,} \label{eq:rate}
\end{equation}
where $\bar{n}_\xi$ is the mean number of motional quanta. Equation~\ref{eq:rate} already provides insight into the dependence of the heating rate on the trap frequency: For a metal, we expect the dissipated power
$\bar{P}_\mathrm{m}$ to be proportional to $\omega_\xi^2$ (see Eq.~11
of Ref.~\cite{Kumph_2016}), implying that the electric-field noise from a metal $S_\mathrm{E,m}$ is
independent of frequency and that the heating rate is proportional to
$1/\omega_\xi$.
In contrast, for a dielectric, $\bar{P}_\mathrm{d}\propto \omega_\xi$ (Eq.~20, Ref.~\cite{Kumph_2016}), implying that  $S_\mathrm{E,d}\propto 1/\omega_\xi$ and that the heating rate is proportional to $1/\omega_\xi^2$.

\section{\label{sec:Trap}Experimental setup and its FEA model}\label{Sec:Floating}

We now summarize the details of the room-temperature surface ion trap, depicted in Figure~\ref{fig:1b}.
Constant (DC) voltages applied to one set of electrodes confine a single $^{40}$Ca$^+$ ion along $z$, while radiofrequency (RF) voltages applied to another set of electrodes confine the ion in the $xy$ plane.  The vertical distance $d$ between the ion and the trap surface can be varied between $\SI{50}{\micro\meter}$ and $\SI{100}{\micro\meter}$ by adjusting the voltages on the DC electrodes.  
The ion is trapped above an electrically floating metal wire, which connects this trapping site to a second one with identical electrode geometry $\SI{620}{\micro\meter}$ apart. The ion's motion induces oscillating image charges in the wire, which are conducted through the wire to the other trapping site. The wire thus acts as a bus between the two trapping sites, enabling coupling between two ions in a recent experiment~\cite{an2022coupling}. All DC and RF electrodes are RF grounded to the vacuum chamber through filter circuits. 
Further details on the design, fabrication and operation of the ion trap can be found in Refs.~\cite{an2022coupling,an2018}.

In FEA simulations, we model the trap substrate as fused silica (SiO$_2$) with thickness $\SI{500}{\micro\meter}$, relative permittivity $\varepsilon_\mathrm{r} =  3.8$, and loss tangent $\tan{\delta}_\mathrm{ref} = \SI{1.0e-3}{}$~\footnote{The values for the loss tangent of fused silica are typically around $10^{-4}$ up to values of $10^{-3}$~\cite{Heraeus}. We use $\tan{\delta}_\mathrm{ref}$ as a worst-case estimate.}. The electrodes are modeled as an aluminum layer with $\SI{1}{\micro\meter}$ thickness and conductivity $\sigma_\mathrm{Al} = \SI{3.8E7}{S/m}$~\cite{Giancoli}.
We determine the boundary conditions in the FEA simulation by a Faraday cage of dimensions $\SI{0.95}{\milli\meter}\times\SI{0.5}{\milli\meter}\times\SI{1}{\milli\meter}$ around the ion trap to which the DC and RF electrodes are connected with infinite conductance. This effectively models the effect of the surrounding vacuum chamber and is valid for chamber dimensions larger than the size of the trap chip and ion-surface distance.

\section{\label{sec:Measurements}Measurements}

First, using the parameters above, we confirm our model by evaluating the noise we expect from a geometry in which the dielectric is mostly shielded by an electrode that is electrically floating~\cite{an2022coupling}. 
We determine the heating rate  $\dot{\bar{n}}_{z}$ of a $^{40}$Ca$^+$ ion in the surface trap of Fig.~\ref{fig:1b} from Rabi oscillations on the $\ket{4^2\mathrm{S}_{1/2}}$ to $\ket{3^2\mathrm{D}_{5/2}}$ qubit transition:
First, the ion is Doppler cooled; next, the experiment is paused for a heating time~$t$, during which the electric field noise increases the ion's kinetic energy; finally, a laser pulse resonant with the qubit transition is applied. 
The laser beam is perpendicular to the trap surface and thus probes the temperature of the motional mode along the $z$ axis. We vary the length of the resonant pulse and extract the mean phonon number $\bar{n}_z$ from fits to the Rabi oscillations, where the Debye-Waller contribution to the fit function provides the phonon-number dependence~\cite{Wineland1998,rowej2002transport}. These measurements are repeated for a range of heating times of $t$, allowing us to determine $\dot{\bar{n}}_{z}$. 

Next, we vary the distance $d$ 
between $\SI{60}{\micro\meter}$ and $\SI{100}{\micro\meter}$. At each distance, we extract both $\dot{\bar{n}}_{z}$ and the motional frequency $\omega_{z}$.
The measured values are listed in Tab.~\ref{Tab:measurements}. 
Under the assumption that heating rates are dominated by noise from dielectrics and thus proportional to $1/\omega_{z}^2$, we normalize the measured heating rates to a motional frequency of $\omega_{z,\text{norm}} = 2\pi\cdot\SI{1}{\mega\hertz}$ with 
\begin{equation}
\dot{\bar{{n}}}_{z, \textrm{norm}} = \dot{\bar{n}}_{z}\frac{\omega_{z,\mathrm{norm}}^2 }{\omega_{z}^2} \mathrm{.}
\label{eq:norm}
\end{equation} These values are also given in Tab.~\ref{Tab:measurements}. This approach allows us to study the heating rate as a function of a single parameter, $d$; alternately, for each distance $d$, we could have adjusted the DC voltages until we reached our target trap frequency.

\setlength{\tabcolsep}{10pt} 
\renewcommand{\arraystretch}{1.5}
\begin{table}
\begin{tabular}{c c c c}\toprule
$d$ & $\omega_{z}/2\pi$ & $\dot{\bar{n}}_{z}$ &$\dot{\bar{n}}_{z,\mathrm{norm}}$  \\ 
$(\mathrm{\mu m})$ &  (MHz) &  (quanta/ms)& (quanta/ms) \\
\midrule
60 & 1.30 & 331(151) & 559(254) \\ 
\hline 
60 & 1.32 & 393(200) & 685(349) \\ 
\hline 
60 & 1.62 & 214(27) & 562(70) \\  
\hline
70 & 1.15 & 338(31) & 447(41) \\ 
\hline 
70 & 1.15 & 327(30) & 432(40) \\ 
\hline 
80 & 1.37 & 202(63) & 378(118) \\ 
\hline 
100 & 1.0 & 152(17) & 152(17) \\ 
\bottomrule
\end{tabular}
\caption{Trap frequencies $\omega_{z}/2\pi$ and heating rates $\dot{\bar{n}}_{z}$ (both measured and normalized) for distances $d$ between $60$ and $\SI{100}{\micro\meter}$.}\label{Tab:measurements} 
\end{table}

\begin{figure}
\includegraphics[scale=0.49]{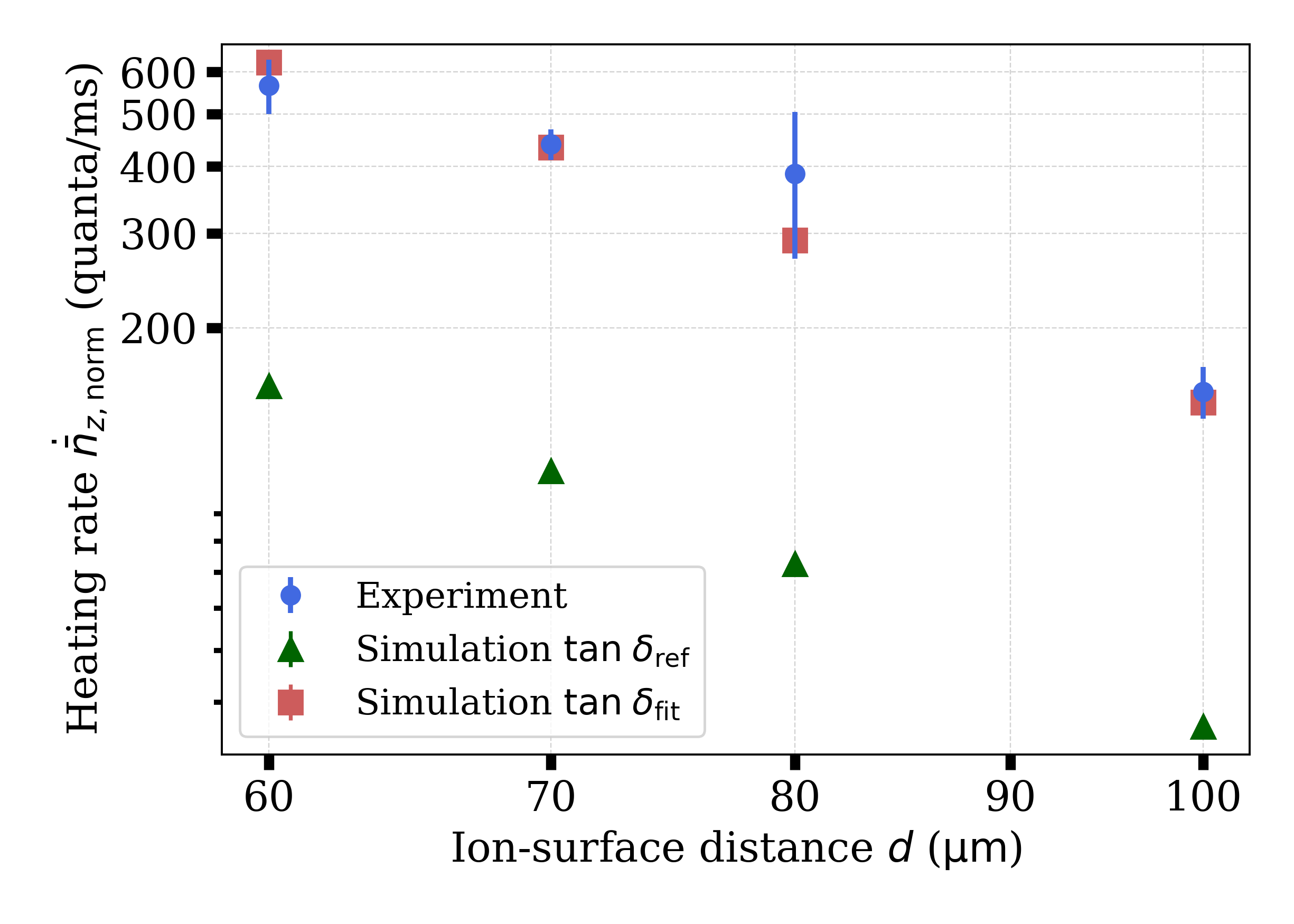}
\caption{Normalized heating rates of the ion's motion along the $z$ axis as a function of the distance $d$ from ion to trap surface, in the trap shown in Fig.~\ref{fig:1b}. All error bars represent one standard deviation. Circles: rates obtained from experiments. The rates correspond to the weighted means of the values listed in Tab.~\ref{Tab:measurements}. Triangles: rates predicted by Eq.~\ref{eq:rate} assuming a loss tangent $\tan{\delta}_\mathrm{ref}$ for the trap substrate.  Relative uncertainties are estimated to be $\SI{1}{\percent}$, too small to be visible. Squares: rates predicted by Eq.~\ref{eq:rate} for a fitted loss tangent. Error bars are determined from the uncertainty of the fitted loss tangent.}
\label{fig:figure3}
\end{figure}

For each value of $d$ and for the fixed frequency $\omega_{z,\mathrm{norm}}$, we simulate the electric-field noise and determine the corresponding heating rates with Eq.~\ref{eq:rate}. 
Both the normalized and simulated heating rates are plotted in Fig.~\ref{fig:figure3}. We observe that both rates decrease with increasing distance $d$. The normalized rates 
range from $152(16)$ to $566(66)\;\mathrm{quanta/ms}$; note that for $d = \SI{60}{\micro\meter}$ and $d = \SI{70}{\micro\meter}$, we have plotted the weighted mean of $\dot{\bar{n}}_{z,\mathrm{norm}}$. The simulated rates with loss tangent $\delta_{\mathrm{ref}}$ are approximately a factor of four smaller. 
The $\SI{1}{\percent}$ relative uncertainties of the simulated rates are estimated from a mesh convergence study, in which we simulate the heating rates for various resolutions of the FEA mesh. 

We 
hypothesize that the discrepancy between experiment and simulation is due to a difference between the loss tangent of the trap substrate used in the experiments and $\tan{\delta}_\mathrm{ref}$.  
The loss tangent is a material property that differs between samples~\cite{Kim} and depends on the specifics of the fabrication process~\cite{Wang} as well as the sample temperature~\cite{MicrowaveElectronics2}; it can vary by more than one order of magnitude for the same material. 
It may also be that residuals of 
solvents applied during and after trap fabrication remained on the trap substrate. Such contaminants are expected to
increase the relative permittivity and loss tangent of the trap substrate, resulting in increased heating rates. 
Under this assumption, we fit the model to the data using the loss tangent as a free parameter common for all distances. The fit yields  $\tan{\delta}_\mathrm{fit} = \SI{4.0(1)e-3}{}$, and a reduced chi-square value $\chi_\nu^2 = 0.89$ demonstrates agreement between the fitted model and the data: Though the loss tangent from the fit is four times larger than the literature value we used~\cite{Note3}, the model accurately captures the observed distance scaling. 

Lastly, we calculate how the metal of the floating electrode contributes to heating rates: We compare our results with an alternate version in which the floating electrode is set to have infinite conductance, such that it does not contribute to the heating rate. We find a contribution of less than $\SI{2.2}{quanta/s}$, six orders of magnitude smaller than $\dot{\bar{n}}_{z}$; that is, $\dot{\bar{n}}_{z}$ is dominated by noise from the dielectric (as we assumed in Eq.~\ref{eq:norm}). 
Nevertheless, although the fluctuations in the electrode's metal have a negligible contribution, the electrode's potential significantly affects the dielectric contribution. When the electrode floats, its potential is allowed to vary; when the electrode is grounded, its potential is fixed, which shields the ion above the trap from the bulk dielectric.
For a grounded electrode, we calculate that $\dot{\bar{n}}_{z}$ is reduced by more than an order of magnitude; for example, for $d=\SI{100}{\micro\meter}$, $\dot{\bar{n}}_{z}$ is reduced by a factor of 200.
This result finds support in measurements reported in Ref.~\cite{an2019distance}, which used a different surface trap of similar electrode geometry fabricated on the same substrate, but with the central electrode held at a fixed potential: the heating rates are more than two orders of magnitude smaller than the values we report in Tab.~\ref{Tab:measurements} for a floating electrode. 

In the context of broader efforts to elucidate the role of materials in quantum information experiments~\cite{deleon2021materials}, we propose the following systematic study: First, the loss tangent of an ion-trap substrate should be measured prior to trap fabrication. Next, a second value of the loss tangent should be extracted following the method described above. A comparison of the two loss-tangent values may shed light on how the fabrication process affects the loss tangent; potentially, this study could be repeated for a range of fabrication conditions.

\begin{figure}
\includegraphics[scale=0.26]{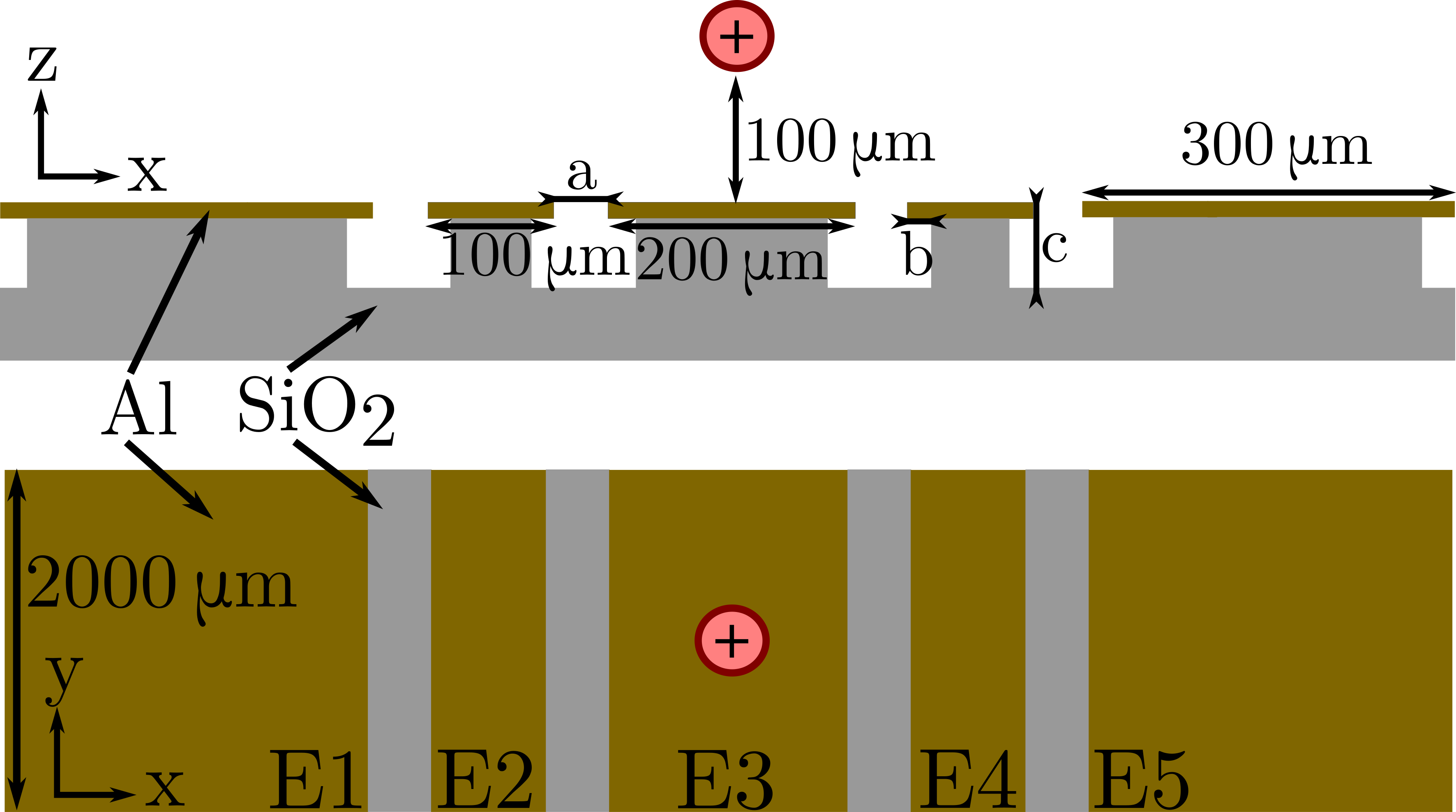}
\caption{A design for a surface ion trap with three variable parameters: the trench width $a$, undercut $b$, and depth $c$. The five electrodes E1$-$E5 are aluminum; the substrate is fused silica. An ion is trapped $\SI{100}{\micro\meter}$ above E3. 
}\label{fig:figure4}
\end{figure}

\floatsetup[figure]{subcapbesideposition=top}
\begin{figure}[!h]
		\sidesubfloat[\label{fig:5a}]{%
			\includegraphics[scale=0.43]{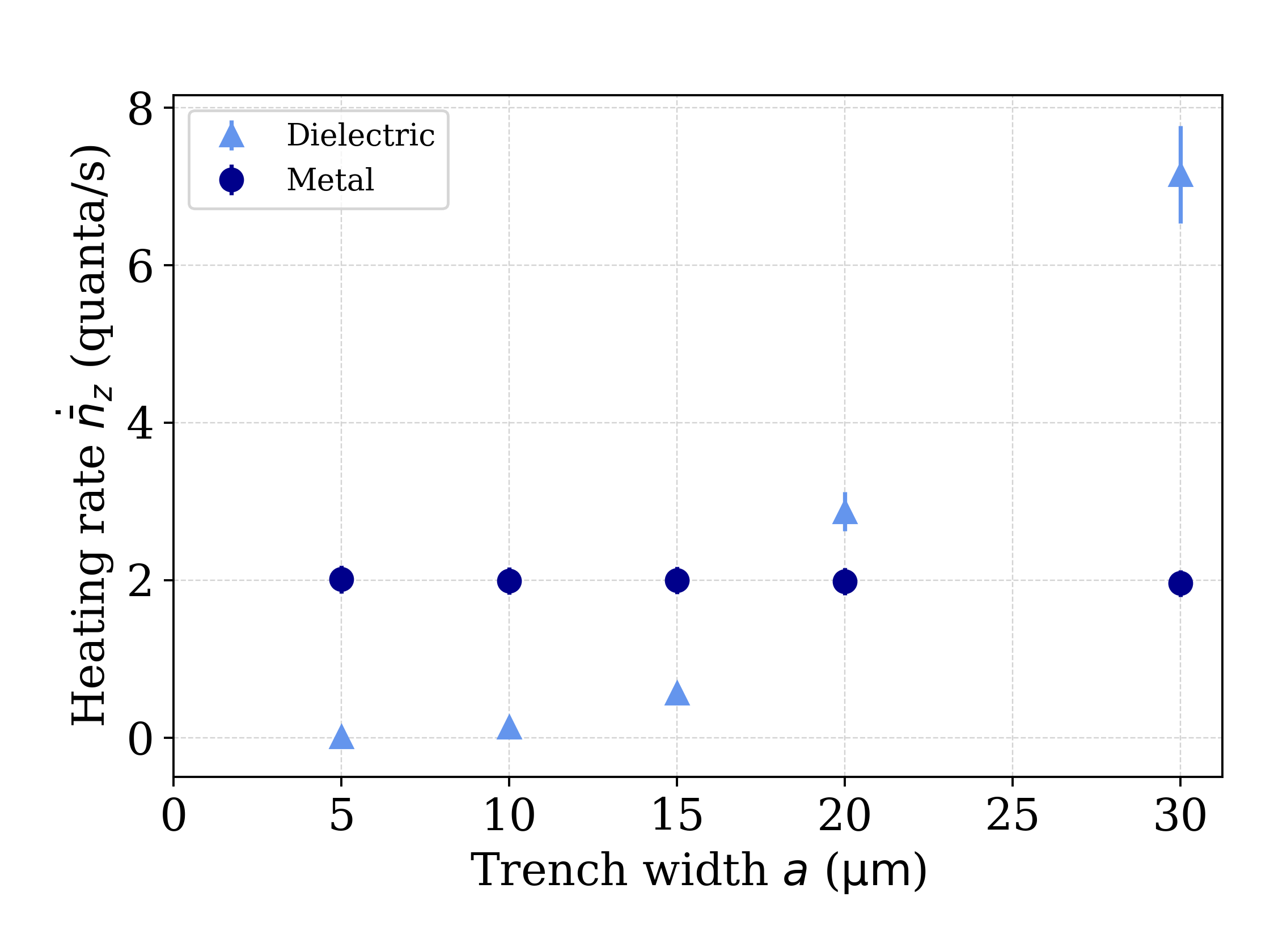}
		}
		\hfill
		\sidesubfloat[\label{fig:5b}]{%
			\includegraphics[scale=0.47]{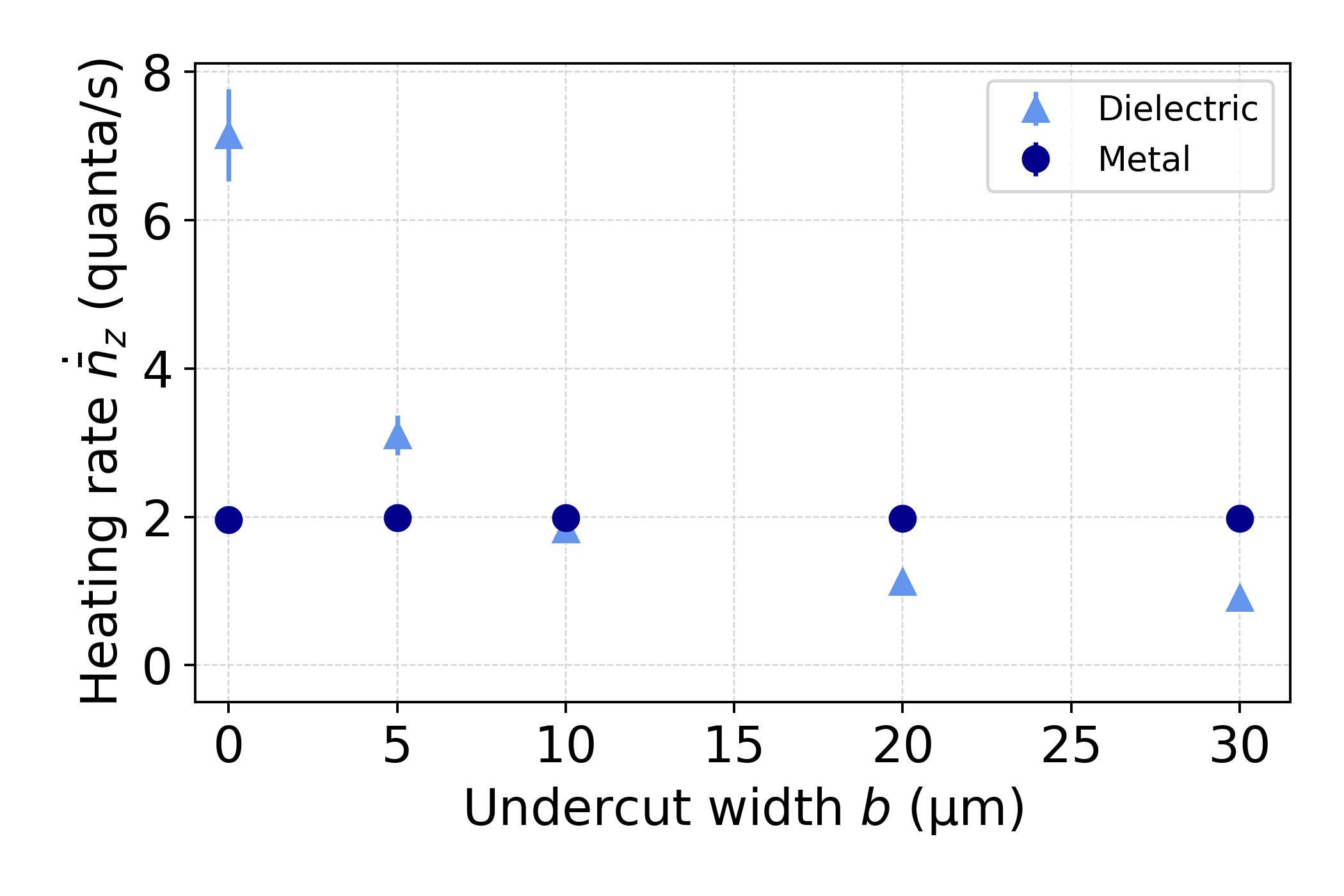}
		}
		\hfill
		\sidesubfloat[\label{fig:5c}]{%
			\includegraphics[scale=0.45]{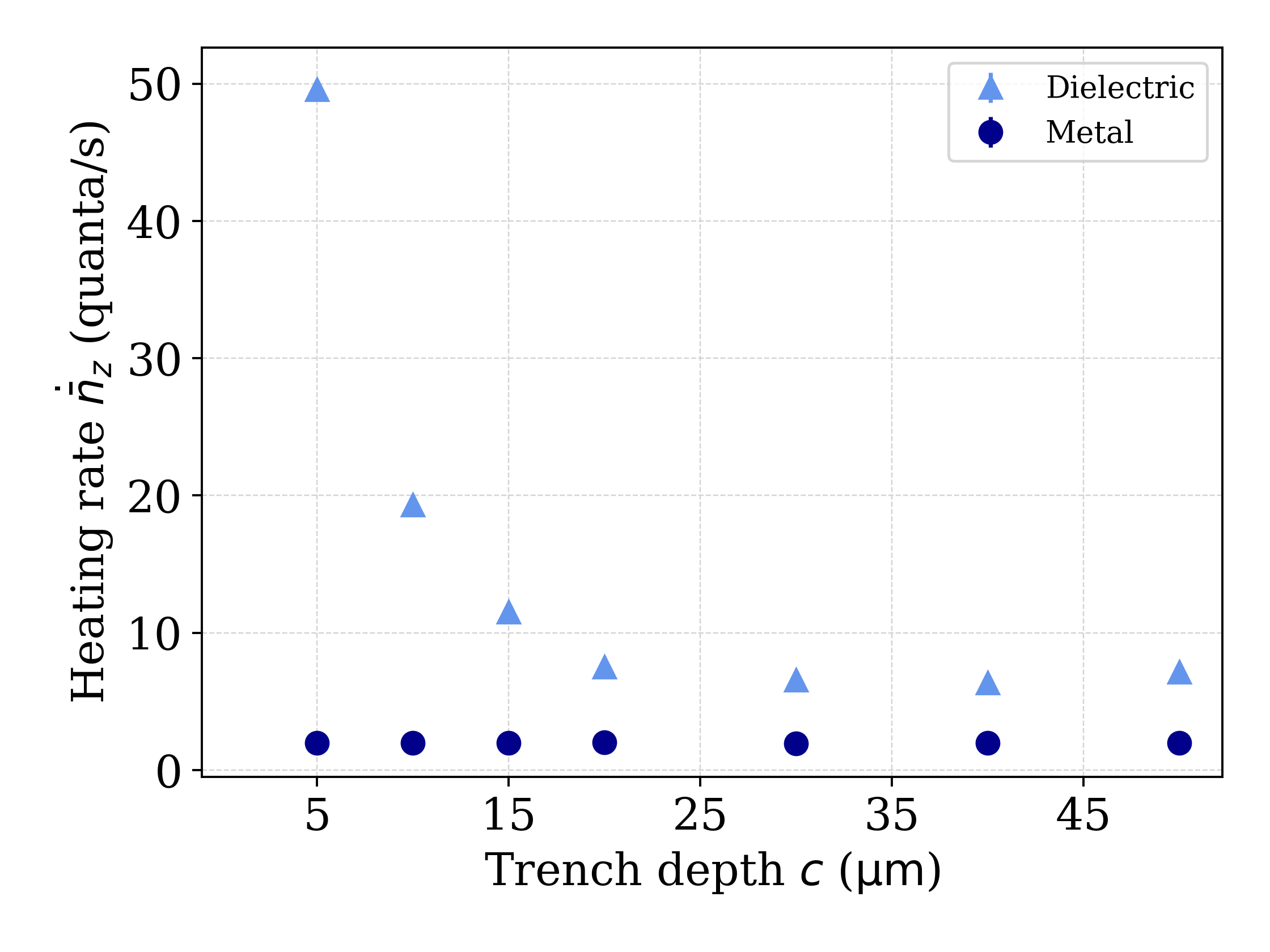}
	}
		\caption{Simulated heating rates along the $z$ axis at a trap frequency $\omega_\mathrm{z} =2\pi\cdot \SI{1}{\mega\hertz}$ and an ion-surface distance $d=\SI{100}{\micro\meter}$, for the trap shown in Fig.~\ref{fig:figure4}. Circles correspond to heating-rate contributions from the metal electrodes. Triangles correspond to heating-rate contributions from the dielectric substrate. 
		(a) The trench width $a$ is varied for fixed values of the undercut width $b=\SI{0}{\micro\meter}$ and trench depth $c = \SI{50}{\micro\meter}$. (b) The undercut width $b$ is varied for for $a=\SI{30}{\micro\meter}$ and $c = \SI{50}{\micro\meter}$. (c) The trench depth $c$ is varied for $a = \SI{30}{\micro\meter}$ and $b = \SI{0}{\micro\meter}$.}
		\label{fig:three graphs}
\end{figure}

\section{\label{sec:Numerics}Numerical studies of a surface ion-trap with variable geometry}

We now apply our noise-simulation method to the task of ion-trap design: We aim to minimize the total electric-field noise seen by an ion in the trap and to examine the relative importance of noise due to metals and noise due to dielectrics. We parameterize the geometry of a surface ion trap and predict the heating rate for each set of parameters, taking into account both dielectric and metal components. In a trap with a geometry similar to the one that we model,
heating rates as low as $\SI{3.8(5)}{quanta/s}$ have been observed  
along the $y$ axis for $d = \SI{100}{\micro\meter}$~\cite{Daniilidis2014}. 

The modeled geometry is illustrated in Fig.~\ref{fig:figure4} and consists of five aluminum electrodes on an SiO$_2$ substrate, a common design for surface ion traps. The electrode thickness is $\SI{1}{\micro\meter}$, and the widths are fixed: $\SI{300}{\micro\meter}$ for the outermost electrodes E1 and E5, $\SI{100}{\micro\meter}$ for the intermediate electrodes E2 and E4, and $\SI{200}{\micro\meter}$ for the central electrode E3. 
Each trench between electrodes is parameterized by its width $a$, depth $c$, and undercut of length $b$ below the electrodes. The parameters $a$, $b$ and $c$ are identical for all four trenches. 
For all simulations, the ion is trapped at a distance $d = \SI{100}{\micro\meter}$ above the electrode E3 of the room-temperature trap, and we use the values $\tan{\delta}_\mathrm{ref}$ and $\sigma_\mathrm{Al}$. The motional modes of the ion align with the axes indicated in Fig.~\ref{fig:figure4}, and the 
with trap frequencies $\omega_\xi= 2\pi\cdot \SI{1}{\mega\hertz}$. As before, a Faraday cage (Fig.~\ref{fig:figure1}) allows us to model the RF grounding of the electrodes in FEA simulations.

\begin{figure}[t]
\includegraphics[scale=0.5]{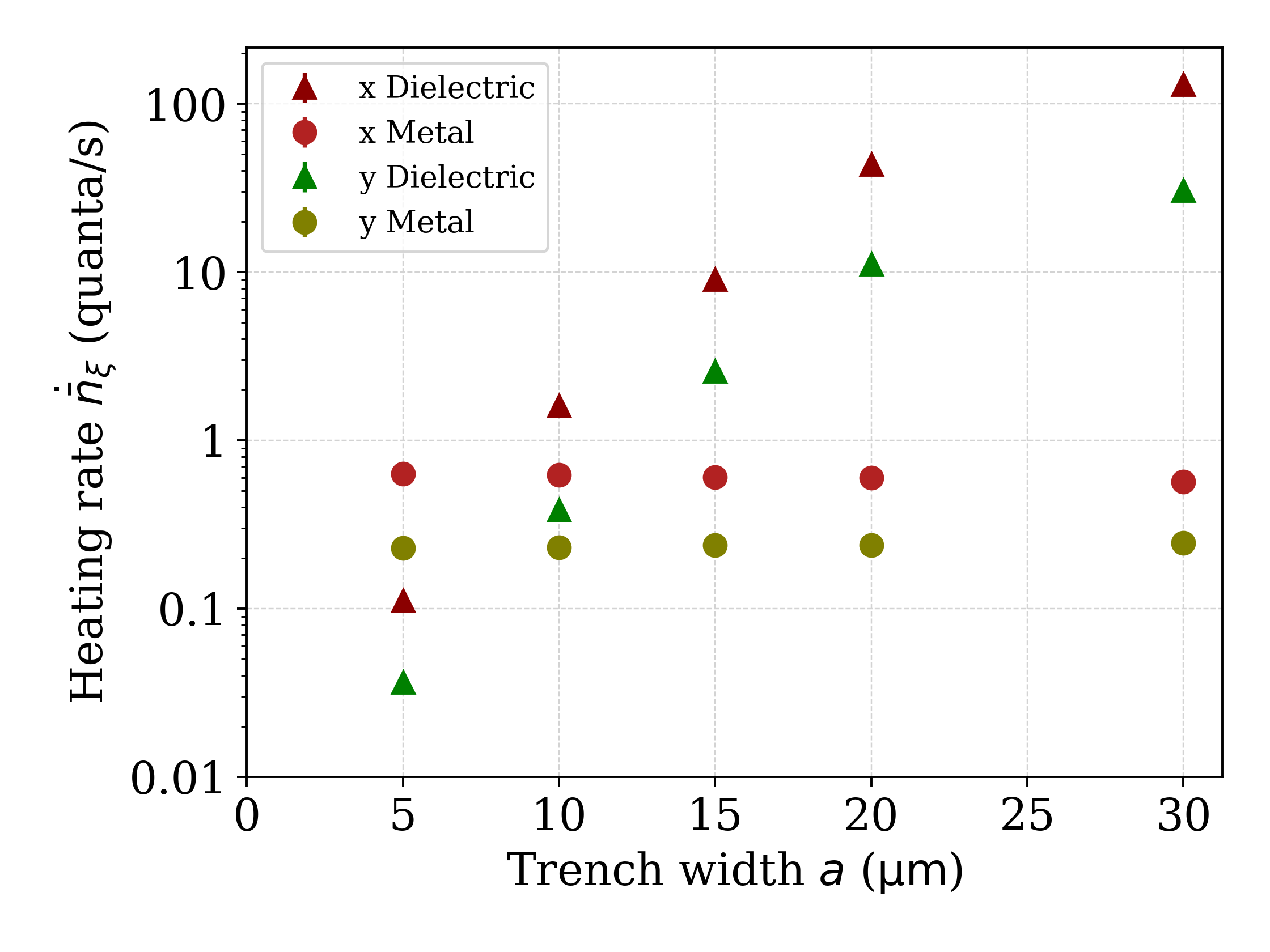}
\caption{Simulated heating rates along the $x$ and $y$ axes under the same conditions as Fig.~\ref{fig:three graphs}a and for trap frequencies $\omega_x = \omega_y =2\pi\cdot \SI{1}{\mega\hertz}$. Circles correspond to  heating-rate contributions from the metal electrodes. Triangles correspond to heating-rate contributions from the dielectric substrate.}\label{fig:figure6}
\end{figure}

Here, we examine heating rates along all three axes, starting with the $z$ axis. We report the results of three studies; in each study, we vary one parameter and calculate the heating rate along $z$.
The first parameter, the trench width $a$, is varied from $\SI{5}{\micro\meter}$ to $\SI{30}{\micro\meter}$. The trench depth is set to $c = \SI{50}{\micro\meter}$, matching the value in Ref.~\cite{Daniilidis2014}. No undercut is considered ($b=\SI{0}{\micro\meter}$). As $a$ is increased, the dielectric contribution $\dot{\bar{n}}_{z,d}$ to the heating rate --- that is, the contribution of $S_\mathrm{E,d}$ to Eq.~\ref{eq:rate} --- increases from $0.010(1)$ to $\SI{7.1(6)}{quanta/s}$ (Fig.~\ref{fig:three graphs}a). In contrast, the contribution of $S_\mathrm{E,m}$ to Eq.~\ref{eq:rate} remains constant 
at $\dot{\bar{n}}_{z,m}=\SI{2.0(2)}{quanta/s}$. 
As explained in Sec.~\ref{Sec:Theory}, we calculate $S_\mathrm{E,d}$ and $S_\mathrm{E,m}$ separately from the fluctuation-dissipation theorem; here, instead of summing them, we plot their separate contributions to the heating rate. 
(Uncertainties are estimated from a mesh convergence study to be $\SI{9}{\percent}$ for all data shown in Figs.~\ref{fig:three graphs} and \ref{fig:figure6}.)
Thus, for small gaps $a\leq \SI{15}{\micro\meter}$, heating due to the metal dominates. 
It is straightforward to interpret this result qualitatively: the ion becomes more exposed to the dielectric when the trenches are widened. In contrast, the important contribution of this predictive method is quantitative information about when the noise is metal-dominated and when it is dielectric-dominated. From this information, we learn in which circumstances it is advantageous to reduce dielectric exposure.

Next, the second parameter, the undercut width $b$,
is varied from $\SI{0}{\micro\meter}$ to $\SI{30}{\micro\meter}$ for $a = \SI{30}{\micro\meter}$ and $c=\SI{50}{\micro\meter}$. The simulated heating rates are plotted in Fig.~\ref{fig:three graphs}b. Here again, the metal contribution remains constant for all values of $b$, and changes in the total heating rate are due to the dielectric contribution, which decreases with increasing values of $b$.
As before, the qualitative interpretation is straightforward: the undercut shields the ion from dielectric-induced noise. Again, the message from simulations is quantitative: for this geometry, temperature and materials, $b =\SI{10}{\micro\meter}$ is the threshold below which the heating rate is dominated by dielectrics.

Finally, the third parameter, the trench depth $c$, is varied from $\SI{10}{\micro\meter}$ to $\SI{50}{\micro\meter}$ for $a = \SI{30}{\micro\meter}$ and $b = \SI{0}{\micro\meter}$, with results shown in Fig.~\ref{fig:three graphs}c. Once more, changes in the heating rate are due solely to the dielectric contribution. The heating rate decreases with increasing trench depth, leveling off for depths greater than $\SI{20}{\micro\meter}$, and we infer that trench depths $c\geq a$ are required to minimize the heating rate. For all trench depths, the dielectric contribution is larger than the metal contribution; this observation is consistent with the fact that $a = \SI{30}{\micro\meter}$ represents a relatively wide trench depth, based on what we have learned from the first study, and that the lack of overhang ($b = \SI{0}{\micro\meter}$) significantly increases the dielectric contribution, based on what we have learned from the second study.

Heating is also present parallel to the trap surface, along the $x$ and $y$ axes. In our last simulation, heating rates along these axes are calculated for the same conditions as in Fig.~\ref{fig:three graphs}a and plotted in Fig.~\ref{fig:figure6}. 
The metal contributions remain constant at $\dot{\bar{n}}_{x,m} = 0.60(5)\;\mathrm{quanta/s} $ and $\dot{\bar{n}}_{y,m} = 0.23(2)\;\mathrm{quanta/s}$; the dielectric contributions $\dot{\bar{n}}_{x,d}$ and $\dot{\bar{n}}_{y,d}$ range over three orders of magnitude. For all values of $a$, $\dot{\bar{n}}_y$ is smaller than $\dot{\bar{n}}_x$ for both noise components. For $a\geq \SI{10}{\micro\meter}$, noise due to the dielectric dominates over $\dot{\bar{n}}_{m}$ for both the $x$ and $y$ axes.
Comparing Figs.~\ref{fig:three graphs}a and~\ref{fig:figure6}, we see that the radial heating rates $\dot{\bar{n}}_x$ and $\dot{\bar{n}}_y$ are larger than $\dot{\bar{n}}_z$ once dielectric noise is dominant, and that dielectric noise starts to dominate along $x$ and $y$ for narrower trench widths. These results point to the importance of simulating heating rates along all three axes, as the optimal geometry may be different for each axis.

In general, we note that for the dielectric noise contribution, the relative strength of the spatial components depends on the orientation of the noise field lines at the ion, which is defined by the electrode layout. For example, for a trench situated directly below the ion, the dielectric noise along $z$ would be dominant. For the metal contribution, the noise depends on both the electrode layout and the locations of the ground connections.
\section{\label{sec:Conclusion}Conclusion}

We have introduced a FEA method for analyzing electric-field noise in ion traps. In contrast to earlier work~\cite{Teller2021}, this method takes into account not only dielectrics but also conducting materials, allowing us to make quantitative predictions of the electric-field noise and ion heating rate in a given trap. The method has been applied to a surface ion trap with an electrically floating metallic wire, yielding agreement between model and experiment when the dielectric loss tangent is a free parameter. The floating wire allowed us to characterize the noise due to the dielectric trap substrate, as the wire propagates the electric-field noise from the substrate to the ion, which then dominates over other noise sources.  Furthermore, we have simulated the heating rate of an ion in a typical surface ion trap along all three trap axes, varying the trench depth, trench width, and undercut width and identifying parameter regimes in which the trap design fails to shield the ion sufficiently from dielectric components.

For a given trap design, this method makes it possible to calculate heating rates that can be directly compared with measured data, which will allow researchers to identify strengths and weaknesses in existing ion traps, including materials choice, fabrication approaches, floating electrodes, and cleaning methods. For example, the heating rates that we determined in our four studies of a trap geometry similar to that of Ref.~\cite{Daniilidis2014} (Figs.~\ref{fig:three graphs} and \ref{fig:figure6}) are comparable to heating rates observed in that trap after in-situ surface cleaning. This suggests that in-situ cleaning can be successful in reducing electric-field noise to a level that is given by the intrinsic losses of the underlying materials. 
More generally, the simulated values for electric-field noise provide a lower bound on what is achievable with a given ion-trap geometry~\cite{Brownnutt2015}. Applying this method as part of an iterative design process will allow ion-trap geometries to be tailored such that the bulk materials contribution to particle heating is below a target value and well understood, offering a useful tool for future quantum simulators and fault-tolerant quantum computers~\cite{deleon2021materials}.

\begin{acknowledgments}
	
This work received funding from the European Union’s Horizon Europe research and innovation programme under Grant Agreement No. 101102140, Project QIA-Phase 1, and ERC-2020-STG 948893; 
the Office of Advanced Scientific Computing Research (ASCR), Office of Science, U.S. Department of Energy, under Contract No. DE-AC02-05CH11231;
and the Austrian Science Fund (FWF) Grants 10.55776/Q4, 10.55776/COE1, and 10.55776/PIN3213524. M.T. acknowledges funding from the European Union's Horizon 2020 research and innovation programme under the Marie Sklodowska-Curie grant agreement 101103143 (2DMultiMems). For open access purposes, the author has applied a CC BY public copyright license to any author-accepted manuscript version arising from this submission.
	
\textit{Note added:} While preparing this manuscript, we became aware of Ref.~\cite{wang2026mitigation}, which addresses a similar question.

\end{acknowledgments}

\bibliography{cqed}

\end{document}